\documentclass[12pt,preprint]{aastex}

\slugcomment{Draft of \today}

\shorttitle{FORCAST}
\shortauthors{Herter et al.}

\begin{document}

\title{First Science Observations with SOFIA/FORCAST: The FORCAST Mid-infrared Camera}

\author{T. L. Herter\altaffilmark{1}, 
J. D. Adams\altaffilmark{1},
J. M. De Buizer\altaffilmark{2},
G. E. Gull\altaffilmark{1}, 
J. Schoenwald\altaffilmark{1}, 
C. P. Henderson\altaffilmark{1},  
L. D. Keller\altaffilmark{3}, 
T. Nikola\altaffilmark{1},
G. Stacey\altaffilmark{1},
W. D. Vacca\altaffilmark{2}  
}

\altaffiltext{1}{Astronomy Department, 202 Space Sciences Building, Cornell University, Ithaca, NY 14853-6801, USA}
\altaffiltext{2}{Universities Space Research Association, NASA Ames Research Center, MS 211-3, Moffett Field, CA 94035, USA}
\altaffiltext{3}{Department of Physics, Ithaca College, Ithaca, NY 14850, USA}


\begin{abstract}
The Stratospheric Observatory for Infrared Astronomy (SOFIA) completed its first light flight in May of 2010 using the facility mid-infrared instrument FORCAST. Since then, FORCAST has successfully completed thirteen science flights on SOFIA. In this paper we describe the design, operation and performance of FORCAST as it relates to the initial three Short Science flights.  FORCAST was able to achieve near diffraction-limited images for $\lambda >$ 30\micron \ allowing unique science results from the start with SOFIA.  We also describe ongoing and future modifications that will improve overall capabilities and performance of FORCAST.

\end{abstract}



\section{Introduction}
The Faint Object infraRed CAmera for the SOFIA Telescope (FORCAST) is a wide-field camera designed to perform continuum and narrow band imaging in the infrared from 5-40\micron. FORCAST is a facility instrument on the Stratospheric Observatory For Infrared Astronomy (SOFIA).  The emission in this part of the spectrum arises mainly from dust heated via UV and optical photons in regions of star formation, evolved stars, and active galactic nuclei.  FORCAST achieves the highest spatial resolution possible with SOFIA since it covers the expected transition region from jitter/seeing dominated to diffraction limited performance.  Originally scheduled for first light before Spitzer, FORCAST is significantly less sensitive than Spitzer (or Herschel) due to emission from the atmosphere and telescope.  However, FORCAST still has utility for complementary observations to Spitzer and Herschel delivering up to three times better spatial resolution than Spitzer and providing supplemental wavelength coverage to both Spitzer and Herschel.  In addition, saturation was an issue for Spitzer for a number nearby high mass star forming environments offering opportunities for complementary observations.

First light with SOFIA was achieved with FORCAST on 26 May 2010 (Gehrz et al. 2011).  This was followed-up by two observatory characterization flights (OCF2 and OCF3) in November 2011 and three Short Science flights in December 2011.  A Basic Science program of ten flights with FORCAST were flown in May-June 2012 in support of the general community.  The Short Science flights offered an opportunity for the FORCAST team to obtain science data while informally commissioning a number of operating modes. This was a particularly challenging time since the telescope system was being tuned and the Mission Controls and Communication System (MCCS) which allows instrument communication with the observatory (for instance to set-up and control telescope movement) was in an early stage of development.  SOFIA is a very complex system which must point and maintain arc-second performance in a changing atmospheric environment.  However, despite our modest expectations, near diffraction-limited-performance was achieved at the longest wavelengths of FORCAST at first light.  As a result the Short Science program was successful in achieving its goal and the letters in this issue report on some of these observations.   Short  and Basic Science observations by the German Receiver for Astronomy at Terahertz Frequencies (GREAT, see Heyminck et al. 2008, 2009; G{\"u}sten, R., et al.) will be reported elsewhere.

The basic design of FORCAST is discussed in section two.  Section three covers the imaging performance and sensitivity of FORCAST while section four discusses data reduction.  Section five outlines some of the changes and upgrades FORCAST will undergo prior to instrument commissioning and delivery in mid 2012.

\section{Instrument Description}

The Faint Object infraRed CAmera for the SOFIA Telescope (FORCAST) is a dual-channel mid-infrared camera and spectrograph sensitive from 5--40~$\mu$m built by Cornell University (Adams et al. 2010). Each channel consists of a 256$\times$256 pixel array. FORCAST has slightly different magnifications in the x and y directions of the array resulting in a 3.4$\arcmin$$\times$3.2$\arcmin$ instantaneous field-of-view\footnote{The detectors may be upgraded for Cycle 1 to 1024$\times$1024 pixel arrays, but if implemented, the field-of-view will remain unchanged. See section 5.}, which after post-processing correction yields square pixels of 0$\farcs$768.  The orientation of this field on the sky depends the field rotation at the time of the observation.

The Short Wave Camera (SWC) uses a Si:As blocked-impurity band (BIB) array that is optimized for observing at 5--25~$\mu$m, while the Long Wave Camera (LWC) Si:Sb BIB array is optimized for the 25--40~$\mu$m range. Observations can be made through either of the two channels individually or, by use of a dichroic mirror internal to FORCAST, with both channels simultaneously.

The FORCAST instrument is composed of two cryogenically cooled cameras of functionally identical design. Light enters
the dewar through a 7.6~cm (3.0~in) diameter window and cold stop and is focused at the field stop, where a six position aperture wheel is located. The wheel holds the imaging field stop and a collection of field masks for instrument characterization. In the future it will also hold the slits used for spectroscopy (see Section 5). The light then passes to the collimator mirror (an off-axis hyperboloid) before striking the first fold mirror, which redirects the light into the liquid helium cooled portion of the cryostat. The incoming beam then reaches a slide, which includes an open position, a mirror, and two dichroics (optimized for different wavelengths). The open position of the slide passes the beam to a second fold mirror, which sends the beam to the LWC, while the mirror position redirects the light to the SWC. The coated silicon dichroics reflect light below 26~$\mu$m to the SWC and pass light from 26--40~$\mu$m to the LWC. The light then passes through a Lyot stop where two filter wheels of six positions each are located, allowing combinations of up to 10 separate filters per channel. Well-characterized, off-the-shelf filters can be used, since a standard 25~mm diameter is used. Finally, the incoming beam enters the camera block and passes through the camera optics. These two-element reflecting systems are composed of an off-axis hyperboloid mirror and an off-axis ellipsoid mirror that focuses the light onto the focal plane array. Also included is an insertable pupil viewer that images the Lyot stop onto the arrays to facilitate alignment of the collimator mirror with the telescope optical axis.

\begin{deluxetable}{lcc}
\tablecaption{FORCAST Short and Basic Science Filter Properties}
\tablewidth{0pt}
\tablehead{
Channel & \colhead{$\lambda$$_{eff}$ ($\mu$m)} & \colhead{$\Delta$$\lambda$\tablenotemark{a} ($\mu$m)}
}

\startdata

SWC & 5.4  & 0.16 \\
    & 6.4  & 0.14 \\
    & 6.6  & 0.24 \\
    & 7.7  & 0.47 \\
    & 8.6  & 0.21 \\
    & 11.1 & 0.95 \\
    & 11.3 & 0.24 \\
    & 19.7 & 5.5 \\
    & 24.2 & 2.9 \\
LWC & 31.5 & 5.7 \\
    & 33.6 & 1.9  \\
    & 34.8 & 3.8 \\
    & 37.1 & 3.3 
\enddata

\tablenotetext{a}{These are based on the half-power points of the filter profiles shown in Figure 1.}

\end{deluxetable}

The SWC and LWC arrays were selected to optimize performance across the 5--40~$\mu$m bandpass. Both arrays have a quantum efficiency (QE) greater than 25\% over most of their used range. Even at airborne altitudes the thermal background is large, requiring the detectors be read out and reset  continuously. The cameras are operated at frame rates between 30 and 300 Hz in either high or low capacitance modes (with full well depths of 1.8$\times$10$^7$ and 1.9$\times$10$^6$ e respectively) depending upon the sky background and source brightness.  The frame rate is chosen to optimize the signal-to-noise ratio and fixes the integrated background level in the array for all observations.  This has the added benefit of minimizing non-linearity corrections since all data is taken at the same well depth.    During data collection frames are coadded to achieved integration times of 5 to 30 seconds.

In dual channel mode, a dichroic is used to split the incoming beam into both the SWC and LWC. Any one of the LWC filters can be used simultaneously with any of the filters in the SWC,  however the presence of the dichroic reduces the overall throughput in both channels. The throughput is filter dependent, however in broad terms, the throughput in dual channel mode relative to the single channel mode is $\sim$60\% from 5--10$\mu$m, $\sim$85\% from 11--25$\mu$m, and $\sim$40\% from 25--40$\mu$m.

Most filters in the SWC are standard Optical Coating Laboratory, Incorporated (OCLI) thin-film interference filters. These filters are stacked with blocking filters to prevent light leaks. The 24.2, 31.5, 33.6, 34.7, and 37.1~$\mu$m filters are LakeShore custom double half-wave (three mesh) filters. The 31.5~$\mu$m filter is a thin film interference filter. The 37.1 and 24.2~$\mu$m filters used were found to have significant blue light leaks. The 37.1~$\mu$m filter light leaks were mitigated by using the dichroic as a blocking filter for wavelengths less than 26~$\mu$m. The 24~$\mu$m filter could not be used with the dichroic and will likely be replaced for Cycle 1 by a similarly designed University of Reading filter, or paired with a diamond scattering blocking filter to provide improved blue-light rejection. In either case, special care had to be taken to provide proper color corrections for objects observed with this filter.

\begin{figure}
\plotone{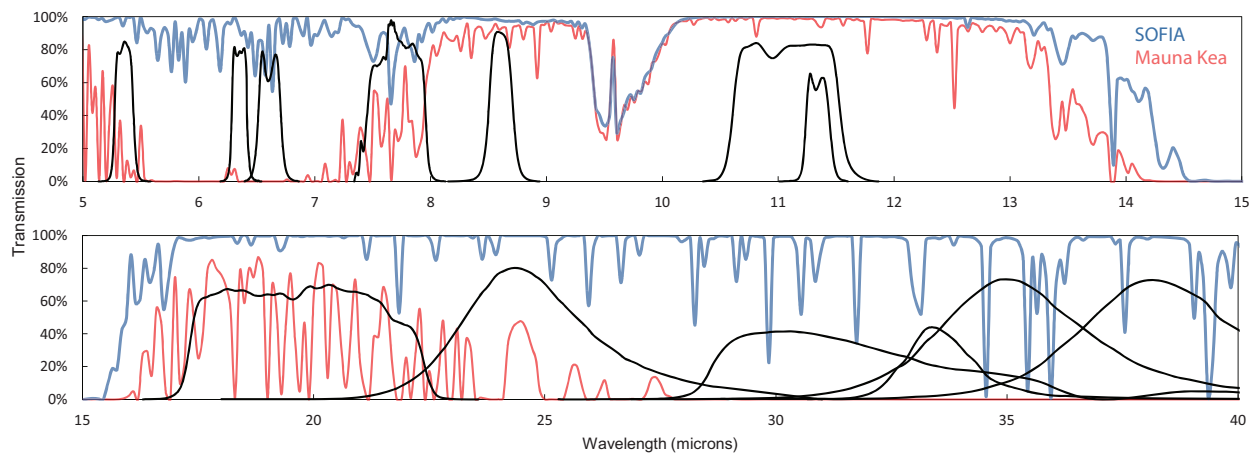}
\caption{Comparison between the atmospheric transmission as observed from a typical SOFIA flight (blue) and a typical night at a ground-based observatory on Mauna Kea (red). Also shown are the transmission profiles for the FORCAST filters used during short and basic science. The SOFIA transmission assumes an altitude of 41000~ft, 7.3 microns of precipitable water vapor, and a telescope zenith angle of 45$\arcdeg$. The Mauna Kea transmission assumes an altitude of 13796~ft and 3.4~mm of precipitable water vapor, and a telescope zenith angle of 45$\arcdeg$. The FORCAST filter profiles are the raw filter transmissions, not accounting for atmospheric, telescopic, and instrumental throughputs.}
\label{fig:images}
\end{figure}

The central wavelengths and bandwidths of each of the filters used in Short and Basic Science are given in Table 1. Figure 1 shows the filter transmission profiles over-plotted on an ATRAN model of the atmospheric transmission (Lord 1992) at typical flight altitude of 41000~ft.

\subsection{Data Acquisition}

Although the altitudes at which SOFIA operates place it well above most of the water vapor in the atmosphere, the observations are still background limited.  Unlike a space-based telescope, the warm atmosphere and warm telescope provide background emission that limit the sensitivity floor. Furthermore, the typical source signal is about $10^{-4}$ the background.  Therefore, as with ground-based IR observatories and unlike space-based observatories, chopping and nodding are essential for obtaining scientifically useful mid-infrared data on SOFIA.

Data are acquired in a standard chop-nod sequence with a typical chop-nod cycle with 30 seconds of integration split between the two chop beams for each nod position.  Two chop-nod sequences are defined: C2N and C2NC2.  For C2N, the chop is symmetric about the optical axis and the source appears on-chip in at least one chop beam of each nod.  For sources of small spatial extent all four chop-nod beams can be placed on chip so all beams can be coadded to achieve a full 60 seconds of integration on source.  A variant of this is a ``matched'' chop-nod which places the plus beam of one nod on top of the minus beam so that no alignment and post-addition of the beams is necessary.  For large sources it is necessary to chop off chip to avoid self-subtraction.  But a large chop introduces coma into the PSF (about 2\arcsec \  per 1\arcmin \  of off-axis tip).  The C2NC2 mode preserves image quality by using an asymmetric chop to place the source on the optical axis for one of the chop beams.  However, this means that the second nod position must be completely off the source (the other chop position has too much coma to be usable) resulting in only 15 out of every 60 seconds spent collecting photons from the source. In both modes the chop-nod cycle is repeated every sixty seconds returning to a slightly different position allowing for removal of bad pixels when the images are combined.  

\section{SOFIA/FORCAST Performance in the Mid-IR}

The telescope optics are designed to provide 1.1\arcsec\ FWHM images on-axis at 0.6\,$\mu$m\ with diffraction-limited performance at wavelengths longer than 15\,$\mu$m.   Furthermore observations from the tarmac with SOFIA show that the combined telescope-FORCAST optical train delivers diffraction limited images down to $\lambda \approx$ 10\micron.  However, the telescope is subject to various vibrations as well as variable windloads in flight, which affect the telescope pointing stability and hence the delivered image quality. SOFIA has active and passive damping systems designed to mitigate these effects. During the last observatory characterization flight (OCF3), after limited tuning of the damping systems, the telescope produced an image quality of 2.8\arcsec\ FWHM at 19.7\,$\mu$m\ with an RMS pointing stability of 1.4$\arcsec$. It should be pointed out that, while these values are large enough to affect image quality at the shorter end of the 0.3--1600~$\mu$m wavelength range of SOFIA, the FORCAST results imply that the observatory was performing at the diffraction-limit for $\lambda$$>$40$\mu$m (i.e. over the large majority of the SOFIA operating wavelength range) just after First Light.

The average image quality obtained during OCF3 through each of the FORCAST filters is shown in Figure 2.   As seen in this figure, telescope pointing stability has a significant impact on the observed image quality in the mid-infrared. The plot shows the contributions of diffraction and the measured RMS telescope pointing stability. The reasons for the deviation from the expected values at shorter wavelengths ($<$8~$\mu$m) is unclear, however this trend continued through all science flights and the cause is still being investigated.  Factors affecting image quality and plans for improvement are discussed in Young et al (2011).

\begin{figure}
\plotone{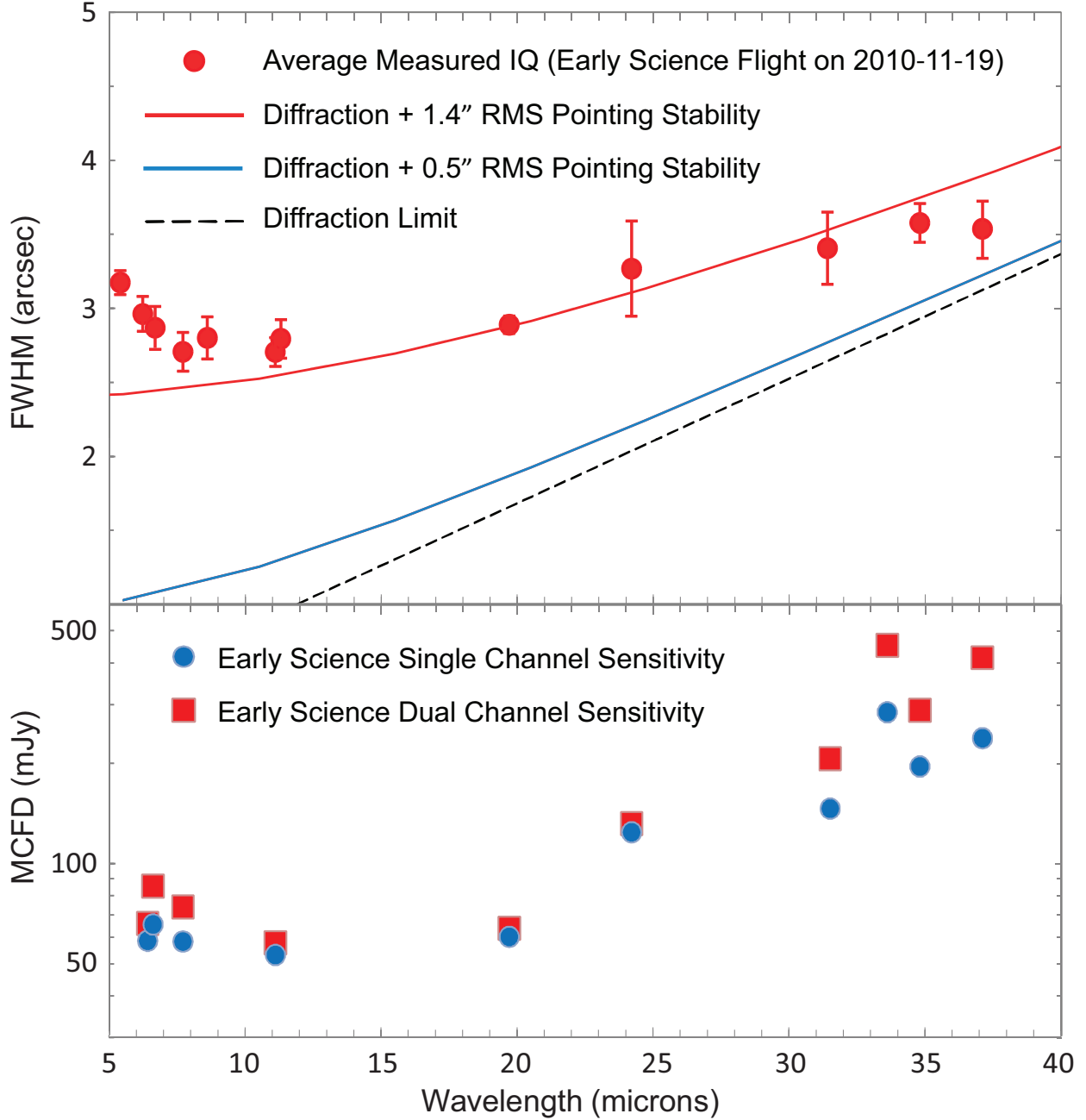}
\caption{Measured mid-infrared image quality (IQ) and sensitivities. (Top) Average FWHM measurements (red dots) during an the last observatory characterization flight (OCF3) as a function of wavelength, with their 1-$\sigma$ standard deviations. Also plotted are the theoretical diffraction limit for a 2.5-m telescope (dashed line) and the limit imposed by the pointing stability of the telescope during the science observations (red line) and the targeted value in the future (blue line). (Bottom) Sensitivity measurements as a function of wavelength for FORCAST during the short and basic science period for a continuum point source. The values reported are the minimum detectable continuum fluxes (MDCF) for a S/N of 4 detection in 900~s at an altitude of 41000~ft and a water vapor overburden of 7~$\mu$m. The red squares indicate the MDCF values in Single Channel mode while the blue circles are for Dual Channel mode using the dichroic. }
\label{fig:images}
\end{figure}

Figure 2 presents the FORCAST imaging sensitivities from observations from OCF3 for a continuum point source at the effective wavelengths of each of the filters. The Minimum Detectable Continuum Flux (MDCF; 80\% enclosed energy) in mJy for a S/N = 4 detection in 900 seconds is plotted versus wavelength for both single and dual channel modes. Accounting for the actual in-flight image quality and background emission, the achieved sensitivity is typically within 30\% of that predicted from instrument performance models and lab measurements.  The sensitivity values in Figure 2 are based on actual OCF3 data taken at 41000 ft, with an assumed precipitable water vapor value of 7~$\mu$m.  Water vapor will generally vary with aircraft altitude, telescope elevation, and height of the tropopause (which itself can vary with season, earth-latitude of the aircraft, and underlying topography).  In general, however, the atmospheric transmission and hence FORCAST response was relatively stable (see next section).

\section{Data Reduction and Calibration}
There are a number of subtleties involved in the reduction of FORCAST data.  For the present purposes we give only a quick overview;  details are given by Herter et al. (2012).  As described earlier, a given observation sequence on an object consists of a set of spatially-dithered, chop-nod positions.  These are processed in the FORCAST pipeline as the following steps:  ``droop'' correction, chop subtraction, nod subtraction, channel subtraction, non-linearity correction, calibration, and averaging.  Droop is a drop in the output signal caused by the presence of the signal itself and is most noticeable in the presence of a bright source which causes a signal offset in nearby channels.  After chop and nod subtraction, ``channel subtraction'' removes correlated noise introduced by bad pixels that ring through the array.  A global non-linearity correction is applied based on the well depth of raw data.  Calibration uses an average flat-spectrum  ($\nu F_{ \nu} = $ constant) source response (electrons/s/mJy) derived from a network of standard stars and solar system objects observed during the Short Science and Basic Science flights (Herter et al. 2012).   For the typical red sources observed with FORCAST, the color corrections are small ($\lesssim 2\%$) and are applied on a case-by-case basis.

Once data are calibrated the next step is to combine the dithered images.  Using a source as a positional reference, images are distortion-corrected, rotated, aligned, and then averaged together.   Bad pixels are marked and eliminated during this last step.  Overall, the pipeline is very robust eliminating virtually all array artifacts.  Typical artifact residuals are smaller than the random noise component of the image and sensitivity scales as expected when averaging data sets.  Bright sources are somewhat problematic since they cause an additional ringing in the array in a way that is not fully removed by channel subtraction, an effect still being investigated.

Notably there is no flat fielding step in the pipeline.  Various attempts to generate a reliable flat field failed.  Three examples are: 1) imaging an extended, uniform source onto the pupil of FORCAST, 2) using raw (unsubtracted) images from the data itself and 3) mapping point sources across the focal plane in the lab.  All of these flats failed to consistently reduce the photometric scatter of in-flight data (Herter et al. 2012).  Because of time limitations it was not possible to map a source across the focal plane in-flight.  Work is ongoing to resolve this issue.  However, the observed 10-20\% peak-to-peak global response variations over the field-of-view are flattened considerably by dithering.

\begin{figure}
\plotone{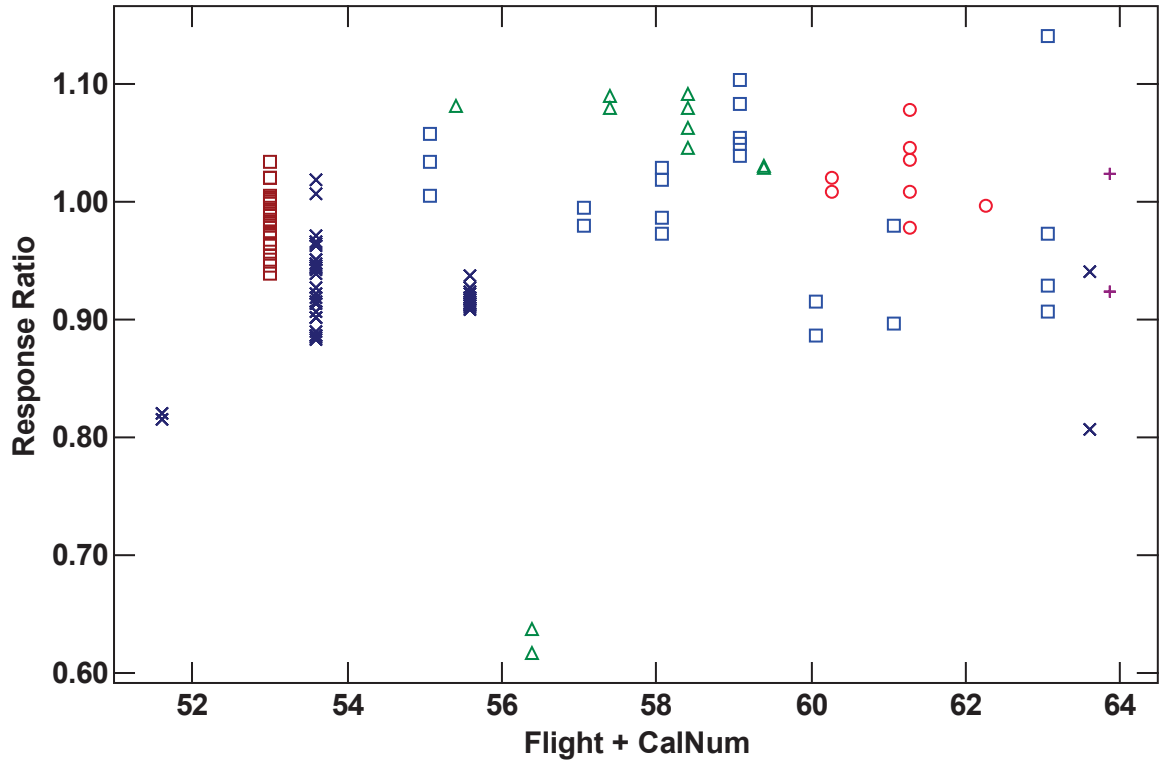}
\caption{FORCAST calibrator response relative to average response at 37\micron \ for Short Science and Basic Science flights.  Each symbol type represents a different calibrator.  The horizontal axis is the flight number plus a fractional offset to separate calibrators.  Flights 51-54 are Short Science and 55-64 are Basic Science.  Only measurements with signal-to-noise ratio greater than thirty and no known problems such as chopper failure or synchronization issues are included.  The standard deviation excluding flight 56 is approximately 6\%.  No corrections have been applied for aircraft altitude, telescope elevation or changing water vapor content.}
\label{fig:images}
\end{figure}

Figure 3 displays the measured responses relative to the average response at 37\micron \ for the Short and Basic Science flights for calibrators with a signal-to-noise ratio greater than thirty and no known issues such as chopper failure or synchronization problems. Excluding flight 56 for which all signals were systematically low (for as yet unknown reasons), the standard deviation is 6\%.   Figure 3 is representative (and perhaps a worst case) of the variation seen at other wavelengths.  Figure 3 contains no corrections for flat field, water vapor burden, altitude or elevation.  Roughly, we adopt a peak-to-peak uncertainty of $20\%$ for the overall absolute flux calibration.  

\section{Future Performance in the Mid-IR}

There is a planned upgrade of the FORCAST detectors for Cycle 1. The sensitivity of FORCAST is background limited but array artifacts (e.g. droop and channel cross-talk as discussed in section 4)  can affect data quality, particularly for extended sources.  These new detectors promise to have better noise qualities and potentially higher quantum efficiencies, and it is hoped that there will be a marked improvement in sensitivity.  In addition, we estimate the total emissivity (dewar window, telescope and atmosphere) in-flight to be 30-40\% which is about a factor of two higher than expected.  We are investigating the options of actively cooling the dewar window and resizing the internal Lyot stop to decrease the background emission.    

Filters in the mid-infrared, especially in the 25-40~$\mu$m range have typically poor transmission. New filter materials and technologies are likely to lead to improvements in system throughput and sensitivity. Furthermore, we have not yet fully optimized the chop-nod efficiencies for all the filters and observing modes of FORCAST. Chop and nod duty cycles are likely to improve with improved settle times of the telescope, and finding optimal chop frequencies at each wavelength. Software upgrades will facilitate operational efficiencies as well.

Finally, FORCAST is presently being outfitted with a suite of grisms available for Cycle 1 that provide low resolution (R$\sim$200) long slit spectroscopy with coverage throughout most of the 5--40~$\mu$m wavelength range of FORCAST  (Keller et al. 2010, Deen et al. 2008). There will also be a cross dispersed spectroscopy mode that will provide medium resolution (R$\sim$800-1200) spectroscopy from 5--14~$\mu$m. Two grisms are situated in each SWC filter wheel and two grisms are mounted in one of the LWC filter wheels, so that they have minimum impact on the imaging capabilities of the instrument. The grisms are blazed, diffraction gratings used in transmission and stacked with blocking filters to prevent order contamination. 

\section{Concluding Remarks}

FORCAST has completed three observatory characterization flights and thirteen science flights with SOFIA achieving near diffraction limited in-flight performance.  These early science flights with FORCAST show that SOFIA is close to meeting its original goal of diffraction limited performance down to 15\micron \ and serve as the start of science operations with SOFIA.

\acknowledgments
We thank R. Grashuis, S. Adams, H. Jakob, A. Reinacher, and U. Lampater for their SOFIA telescope engineering and operations support. We also thank
the SOFIA flight crews and mission operations team (A. Meyer, N. McKown, C. Kaminski) for their SOFIA flight planning and flight support.  We also wish to recognize Lea Hirsch and Jason Wang, Cornell undergraduates would joined the FORCAST team to participate in the basic flights and tirelessly sifted through the data set to help with calibration.  This work is based on observations made with the NASA/DLR Stratospheric Observatory for Infrared Astronomy (SOFIA).  SOFIA science mission operations are conducted jointly by the Universities Space Research Association, Inc. (USRA), under NASA contract NAS2-97001, and the Deutsches SOFIA Institut (DSI) under DLR contract 50 OK 0901.  Financial support for FORCAST was provided by NASA through award 8500-98-014  issued by USRA.


\end{document}